\documentclass[11pt, a4paper, twoside, openright]{memoir}
\usepackage[square,comma,numbers,sort&compress]{natbib}
\usepackage{setspace}
\usepackage{graphicx,amsmath,bbm} 
\usepackage[breaklinks,bookmarks]{hyperref}
\usepackage{memhfixc}
\makepagestyle{custom}
\makeevenhead{custom}{\normalfont\thepage}{}{
M.~H.~Szyma\'nska, J.~Keeling, P.~B.~Littlewood}
\makeoddhead{custom}{\rightmark}{}{\normalfont\thepage}

\DeclareMathOperator{\tr}{Tr}

\DeclareMathOperator{\im}{Im}

\setstocksize{310mm}{210mm}
\settrimmedsize{310mm}{210mm}{*}
\settypeblocksize{*}{1.1\lxvchars}{1.8}
\setlrmargins{20mm}{*}{*}
\setulmargins{*}{*}{1}
\setmarginnotes{20pt}{110pt}{\onelineskip}
\checkandfixthelayout

\changecaptionwidth
\captiontitlefont{\small}
\captionwidth{0.8\textwidth}

\begin{document}

\bibliographystyle{apsrev}

\frontmatter
\pagenumbering{roman}

\pagestyle{empty}
\clearpage
\begin{center}

\large{{\textbf{Non-Equilibrium Bose--Einstein Condensation in
a  Dissipative Environment }}}
\vspace{10ex}

Marzena H. Szyma\'nska,
%$^1$, 
Jonathan Keeling and
%$^2$
Peter B. Littlewood
%$^1$ 
\\
\let\thefootnote\relax\footnotetext{ \hspace{-8mm} 
  Marzena H. Szyma\'nska\\ Department of Physics, University of
  Warwick, CV4 7AL, UK  
  \\
  also at London Centre for Nanotechnology \\
Jonathan Keeling \\ 
  Scottish Universities Physics Alliance,
School of Physics and Astronomy, University of St Andrews, KY16 9SS, UK \\ 
 Peter Littlewood \\ 
 Physical Science and Engineering Division, Argonne National Laboratory, 
 9700 S. Cass.\ Av., Argonne, IL 60439, USA; 
\\ James Franck Institute and Department of Physics,  University of
Chicago, IL 60637, USA;   
\\ Cavendish Laboratory, University of Cambridge, CB3 0HE, UK }
\end{center}
\vspace{10ex}
\noindent\textbf{Abstract:}

\noindent

 Solid state quantum condensates can
 differ from other condensates, such as Helium, ultra-cold atomic
 gases, and \index{superconductor}superconductors, in that the
 condensing \index{quasiparticle}quasiparticles have relatively short
 lifetimes, and so, as for lasers, external \index{pumping}pumping is
 required to maintain a \index{non-equilibrium!steady state}steady
 state. In this chapter we present a \index{non-equilibrium!quantum
   field theory}non-equilibrium \index{path integral}path integral
 approach to condensation in a \index{dissipative}dissipative
 environment and apply it to micro\index{microcavity}cavity
 \index{polariton}polaritons, driven out of equilibrium by coupling to
 multiple baths\index{heat bath}, describing \index{pumping!and
   decay}\index{pumping!and decay bath}pumping and decay. Using this,
 we discuss the relation between
 \index{non-equilibrium!condensate}non-equilibrium
 \index{polariton}polariton condensation, lasing, and equilibrium
 condensation.

\clearpage
\pagestyle{custom} \mainmatter

\renewcommand{\thefigure}{\arabic{figure}}

%\author[]{Marzena H. Szyma\'nska}
%\address{Department of Physics, University of Warwick, CV4 7AL, UK} 

%\author[]{Jonathan Keeling} 
%\address{Scottish Universities Physics Alliance, \\
%School of Physics and Astronomy, University of St Andrews, KY16 9SS, UK} 

%\author[M. H. Szyma\'nska, J. Keeling \& P. B. Littlewood]{Peter B. Littlewood} 
%\address{Physical Science and Engineering Division, Argonne National Laboratory, 
%\\ 9700 S. Cass.\ Av., Argonne, IL 60439, USA; 
%\\ James Franck Institute and Department of Physics, \\ University of Chicago, IL 60637, USA;  
%\\ Cavendish Laboratory, University of Cambridge, CB3 0HE, UK} 

%\index[aindx]{Szymanska, M. H.@Szyma\'nska, M. H.} % or \aindx{Author, F.}
%\index[aindx]{Keeling, J.} % or \aindx{Author, F.}
%\index[aindx]{Littlewood, P. B.} % or \aindx{Author, F.}

\section{Introduction}
\label{szymanska_sec_jntro}

The idea of Bose--Einstein condensation of \index{quasiparticle}quasiparticles in
solid-state structures has a long history, since the early proposals
\cite{keldysh_kopaev_65,blatt_boer_62,moskalenko_62} that \index{exciton}excitons
might form a condensate.  In more recent years, this has led to work
on a variety of systems: \index{exciton}excitons in coupled quantum
wells~\cite{butov_lai_02,butov_gossard_02,snoke_denev_02,butov_04};
\index{exciton}excitons in quantum-Hall \index{bilayer}bilayers~\cite{eisenstein_macdonald_04};
\index{magnon}magnons, both in \index{equilibrium!thermal}thermal equilibrium~\cite{ruegg_cavadinin_03} and
parametrically pumped magnetic insulators~\cite{demokritov_demidov_06}
as well as within superfluid $^3$He\cite{volovik08,bunkov_volovik_10};
and micro\index{microcavity}cavity \index{exciton}exciton-\index{polariton}polaritons.
(For extensive references to experiments see the review \cite{deng_haug_10}).
In almost all these cases, the condensate is, to a greater or lesser
extent, a non-equilibrium \index{non-equilibrium!steady state}steady state, with \index{pumping}pumping compensating for
the finite lifetime of the \index{quasiparticle}quasiparticles, leading to a flux of
particles through the system. Thus, a general question arises: can
Bose--Einstein condensation be realised in a strongly \index{dissipative}dissipative
environment, and if so how would it relate to and differ from
equilibrium BEC and the laser?

To address these questions, this chapter discusses a field theoretical
approach to modelling quantum condensates that are driven out of
equilibrium by a flux of particles through the system. We illustrate
the technique using an example of micro\index{microcavity}cavity \index{polariton}polaritons. Their
part-light nature leads to a rather short lifetime that may
nonetheless be long enough to have \index{polariton}polaritons as well-defined
\index{quasiparticle}quasiparticles (i.e. strong coupling).  Their short lifetimes however
lead to an important role of \index{non-equilibrium!phenomena}non-equilibrium physics. This naturally
provokes questions about the relation to lasing, which occurs in
pumped cavities in the weak coupling limit.  We consider a \index{polariton}polariton
system coupled to \index{pumping!and decay bath}baths which model \index{pumping!and decay}pumping and decay processes. Since
these \index{}baths are not in chemical equilibrium with each other, they
drive a flux of particles.  The Hamiltonian\index{Hamiltonian!with pumping and decay} we use will describe both
a laser (if pumped at high temperatures, as discussed below), and Bose
condensation if treated in \index{equilibrium!thermal}thermal equilibrium, as well as the smooth
transition between them.  As such, the system of micro\index{microcavity}cavity
\index{polariton}polaritons provides a particularly rich playground for studying
\index{coherence}coherence in a \index{dissipative}dissipative environment, and exploring the differences
and similarities between condensates and lasers.

\section{Methodology: Modelling the Non-Equilibrium System}
\label{sec:szymanska-modell-non-equil}

For a \index{non-equilibrium!system}non-equilibrium system, the
density of states and its occupation must both be determined
explicitly, as the occupation may be non-thermal.  This means that to
describe the system fully, one needs at least two \index{Green
  function}Green's functions.  We choose here to work with the
retarded and \index{Keldysh}Keldysh \index{Green function}Green's
functions: $ D^R(\mathbf{r},\mathbf{r}^\prime,t,t^\prime) = -i
\theta(t) \langle [ \hat\psi(\mathbf{r},t),
  \hat\psi^\dagger(\mathbf{r}^\prime,t^\prime)]_-\rangle$,
$D^K(\mathbf{r},\mathbf{r}^\prime,t,t^\prime) = -i \langle [
  \hat\psi(\mathbf{r},t),
  \hat\psi^\dagger(\mathbf{r}^\prime,t^\prime)]_+ \rangle$, where
$[\hat\psi,\hat\psi^\dagger]_{\mp}$ is the commutator
(anti-commutator).  The retarded \index{Green function}Green's
function describes the response following some applied perturbation.
In the frequency domain, $\rho(\mathbf{p},\omega)= 2 \im
[D^R(\mathbf{p},\omega)]$ gives the density of states, while the
\index{Keldysh}Keldysh \index{Green function}Green's function
$D^K(\mathbf{p},\omega) = -i [2 n(\omega) + 1]
\rho(\mathbf{p},\omega)$ accounts for occupation $n(\omega)$.

To determine these \index{Green function}Green's functions, we will
use a \index{path integral}path integral
approach\cite{kamenev_chapter_05}, discussed further below.  Path
integrals naturally allow computation of time-ordered correlation
functions; in order to instead find the retarded and
\index{Keldysh}Keldysh \index{Green function}Green's functions we must
use the \index{Keldysh}Keldysh contour $\mathcal C$.  Points on this
contour are labelled by $(t,\{+,-\})$, where $+,-$ distinguishes the
forward($+$) and backward($-$) branches.  The \index{path
  integral}path integral approach will then give contour ordered
correlations, denoted by $T_{\mathcal C}$, where fields on the $+$
contour always precede those on the $-$ contour, and fields on the $-$
appear in time reversed order.  Then introducing symmetric (classical)
and anti-symmetric (quantum) combinations of these fields
$\psi_{\textrm{cl},\textrm{q}} = \left[\psi(+,t) \pm
  \psi(-,t)\right]/\sqrt{2}$, the \index{Green function}Green's
functions are given by:
\begin{equation}
  \label{eq:szymanska-GreenDef}
  D = 
  \left( 
    \begin{array}{cc}
        D^K & D^R \\ D^A & 0 
    \end{array}
  \right)
  =
  -i
  \left\langle
    T_{\mathcal C} \left(
      \begin{array}{c}
        \psi_{\textrm{cl}}(\mathbf{r},t) \\ \psi_{\textrm{q}}(\mathbf{r},t) 
      \end{array}
    \right)
    \left(
      \psi_{\textrm{cl}}^\dagger(\mathbf{r}^\prime,0),  \psi_{\textrm{q}}^\dagger(\mathbf{r}^\prime,0) 
      \right)
  \right\rangle.
\end{equation}
($D^A$ is the advanced \index{Green function}Green's function, the Hermitian conjugate of
$D^R$). As discussed below, the action in the \index{path integral}path integral involves
the inverse \index{Green function}Green's function: $ D^{-1} = \left[ \left(
\begin{smallmatrix}
{D}^{K} & {D}^{R} \\
{D}^{A} & 0
\end{smallmatrix}
\right)\right]^{-1} 
= \left(
\begin{smallmatrix}
0  &  [{D}^{A}]^{-1}  \\
[{D}^{R}]^{-1}  &  [{D}^{-1}]^{K}
\end{smallmatrix}\right)$, 
  where $[{D}^{-1}]^{K} = -
[{D}^{R}]^{-1} {D}^K [{D}^{A}]^{-1}$.  As an
illustration, for a free field $[{D}_0^{R}]^{-1} =
\hbar\omega - \hbar\omega_p + i \delta$ and
$[{D}_0^{-1}]^{K} = 2 i \delta [2 n_B(\hbar\omega) +
1],$ where $\delta$ is infinitesimal.  The above is for bosonic
fields; the results for fermionic fields are similar, but
commutators and anti-commutators are interchanged in the definitions
of \index{Keldysh}Keldysh and retarded \index{Green function}Green's functions.

\subsection{Polariton System Hamiltonian\index{Hamiltonian!polariton gas}, and Coupling to Baths}
\label{sec:szymanska-model-hamilt-coupl}

To describe the \index{polariton}polariton system we use a model of \index{disorder}disorder
localised \index{exciton}excitons strongly coupled to \index{cavity}cavity
photons\cite{eastham_littlewood_00,keeling_eastham_04,marchetti_keeling_06}.
\index{exciton-exciton}Exciton-\index{exciton}exciton \index{interaction(s)!exciton-exciton}interactions are included in this model by allowing
only zero or one \index{exciton}excitons on a given site, thus describing hard-core
bosons.  This model has several advantages for our aims: Firstly, this
same Hamiltonian has been used to model lasers
\cite{haken_book_70}, allowing us to relate \index{polariton}polariton condensation
to lasing.  Secondly, it is known\cite{keeling_eastham_04} that, in
equilibrium, except at extremely low densities, \index{mean-field theory}mean-field theory
captures the phase diagram of this model rather well. Finally, it
allows one to account straightforwardly for \index{exciton}exciton nonlinearity
within the \index{non-equilibrium!mean-field theory}non-equilibrium \index{mean-field theory}mean-field theory.

To describe a hard-core boson, we introduce two fermionic operators
$\hat{d}^\dagger_j, \hat{c}^\dagger_j$ that create states representing
the presence or absence of an \index{exciton}exciton.  The operator
$\hat{d}^\dagger_j\hat{c}^{\vphantom{\dagger}}_j$ thus creates an \index{exciton}exciton.  In
this notation, the system Hamiltonian is $ \hat{H}_{\mathrm{sys}} =
\sum_j \epsilon_j (\hat{d}^\dagger_j \hat{d}^{\vphantom{\dagger}}_j -
\hat{c}^\dagger_j \hat{c}^{\vphantom{\dagger}}_j) + \sum_\mathbf{p} \hbar\omega_\mathbf{p}
\hat\Psi^\dagger_\mathbf{p} \hat\Psi^{\vphantom{\dagger}}_\mathbf{p} + \sum_{j,\mathbf{p}} g_{j}
(\hat\Psi^\dagger_\mathbf{p} \hat{c}^\dagger_j \hat{d}^{\vphantom{\dagger}}_j +
\mathrm{H.c.}).$ Here $\epsilon_j$ is the \index{exciton}exciton state energy and
$g_j$ is the coupling to photons. The \index{cavity}cavity photon dispersion is
$\hbar \omega_\mathbf{p} = \hbar \omega_0 + \hbar^2 p^2/2 m_{\mathrm{phot}}$.

% FIXME; Merge Hamiltonians
The system is driven out of equilibrium by its coupling to separate
\index{pumping!and decay bath} \index{pumping!and decay}pumping and decay baths, so that the full Hamiltonian is given by\index{Hamiltonian, expression for!polariton system}
$\hat{H} = \hat{H}_{\mathrm{sys}} +
\hat{H}^{\mathrm{pump}}_{\mathrm{bath}} +
\hat{H}^{\mathrm{decay}}_{\mathrm{bath}}$.  The contribution of the
pumping bath is $\hat{H}^{\mathrm{pump}}_{\mathrm{bath}} =
\sum\limits_{j,n} \Gamma_{j,n} (\hat{c}^\dagger_j
  \hat{C}^{\vphantom{\dagger}}_{j,n} + \hat{d}^\dagger_j \hat{D}^{\vphantom{\dagger}}_{j,n}
  + \mathrm{H.c.}) + \sum\limits_{j,n} \nu_{j,n}^\Gamma (
  \hat{D}^\dagger_{j,n} \hat{D}^{\vphantom{\dagger}}_{j,n} -
  \hat{C}^\dagger_{j,n} \hat{C}^{\vphantom{\dagger}}_{j,n})$.  The
fermionic operators $\hat{D}^\dagger_{j,n}, \hat{C}^\dagger_{j,n}$
describe the pumping bath\index{pumping!and decay bath} modes, and $\Gamma_{j,n}$ is the coupling
strength.  Similarly, the contribution of the decay bath is $
\hat{H}^{\mathrm{decay}}_{\mathrm{bath}} = \sum\limits_{\mathbf{p},p_z}
\zeta_{\mathbf{p},p_z} ( \hat\Psi^\dagger_{\mathbf{p}}
  \hat\Xi^{\vphantom{\dagger}}_{\mathbf{p},p_z} + \mathrm{H.c.}) +
\sum\limits_{\mathbf{p},p_z} \hbar\omega_{\mathbf{p},p_z}^\zeta
\hat\Xi^\dagger_{\mathbf{p},p_z} \hat\Xi_{\mathbf{p},p_z}$, with
$\hat\Xi^\dagger_{\mathbf{p},p_z}$ describing bulk photon modes.  Each
confined photon mode $\mathbf{p}$ couples to a separate set of bulk photon
modes with various values $p_z$, corresponding to conservation of
in-plane momentum in the coupling between \index{cavity}cavity and bulk photon
modes.

\subsection{Path-Integral Formulation}
\label{sec:szymanska-path-integral}

Following Ref.~\cite{kamenev_chapter_05}, we construct the
\index{non-equilibrium!quantum field theory}non-equilibrium
\index{generating functional}generating functional $\mathcal{Z}$ as a
coherent state \index{path integral}path integral over
fields\footnote{In keeping with the convention of Ref.
  \cite{kamenev_chapter_05}, we also refer to field amplitudes defined
  at discrete momenta, such as $\Psi_{\mathbf{p}}$, as fields.  We
  note that $\mathcal{Z}$ is necessarily a functional integral, as we
  must account for a continuum of paths taken by $\Psi$, $\Lambda$, C,
  D, $\Xi$ (and their complex conjugates) as functions of the
  continuous time variable $t$.} defined on the closed-time-path
contour, $\mathcal{C}$. For conciseness, we arrange the fermionic
fields into a Nambu vector $\Lambda= (d,c)^{\mathrm T}$. Formally, the
partition function is thus: ${\mathcal Z}= \int
\prod_{\mathbf{p},p_z,j,n} {\mathcal D}[\Psi_{\mathbf{p}}, \Lambda_j,
  C_{j,n}, D_{j,n}, \Xi_{\mathbf{p},p_z} ] e^{iS}, $ where the
action\footnote{When evaluating things we tend to take the continuum
  limit over $\mathbf{p}$, making the partial time-derivative more
  convenient and appropriate \cite{kamenev_chapter_05}.}
$S=\int_{\mathcal{C}} dt \langle
\Psi_{\mathbf{p}}(t)\Lambda_j(t)C_{j,n}(t)D_{j,n}(t)\Xi_{\mathbf{p},pz}(t)
| i \hbar \partial_t - H |
\Psi_{\mathbf{p}}(t)\Lambda_j(t)C_{j,n}(t)D_{j,n}(t)\Xi_{\mathbf{p},pz}(t)
\rangle$, in terms of \index{coherent!state}coherent states specified
by the fields $\Psi_{\mathbf{p}}, \Lambda_j ,C_{j,n}, D_{j,n},
\Xi_{\mathbf{p},p_z}$ on the \index{Keldysh}Keldysh time contour
$\mathcal C$.  In what follows, we consider fields in the classical
and quantum rather than forward and backward basis.

%I DON'T THINK DESIGNATION AS FIELDS IS FORMALLY APPROPRIATE, NOR IS THE USE OF FUNCTIONAL INTEGRAL NOTATION.

\subsection{Treatment of Environment}
\label{sec:szymanska-treatm-envir}

%AGAIN, FORMALLY, THESE DO NOT APPEAR TO BE FIELDS (ARE NOT CONTINUOUSLY PARAMETRIZED IN SPACE)

For the bath Hamiltonian given above, the action $S$ contains
only terms linear or quadratic in the bath fields $ C_{j,n}, D_{j,n},
\Xi_{\mathbf{p},p_z}$ and their conjugates. Thus, the integral over
these fields is Gaussian, and can straightforwardly be evaluated
analytically. For the decay bath\index{pumping!and decay bath} one thus finds:
\begin{equation}
S^{\mathrm{decay}}_{\mathrm{bath}}
= - \iint_{-\infty}^\infty \!\!\! dt dt^{\prime}
\sum_{\mathbf{p}} 
\Psi^{*}_\mathbf{p}(t)\sigma_1^{\textsc{k}}
\sum_{p_z} \zeta_{\mathbf{p},p_z}^2\left[(i\hbar\partial_t -
\hbar\omega^{\zeta}_{\mathbf{p},p_z})\sigma_1^{\textsc{k}}\right]^{-1}\sigma_1^{\textsc{k}}
\Psi_{\mathbf{p}}(t^{\prime}),
\end{equation}
where $\sigma_1^{\textsc{k}}$ is a Pauli matrix in the (\index{Keldysh}Keldysh) space
of quantum and \index{classical field}\index{classical field|seealso{Rayleigh--Jeans}}classical fields.  By definition, the bath has a
large number of modes, and these modes thermalise rapidly compared to
system-bath \index{interaction(s)!system-bath}interactions.  Hence we may take the bath\index{pumping!and decay bath} occupation
functions as fixed, and then allow the system distribution to be
self-consistently determined.  The \index{Green function}Green's function of a free bosonic
field is
$[(i\hbar\partial_t - \hbar\omega^{\zeta}_\mathbf{p})\sigma_1^{\textsc{k}}
]^{-1} = \left(
  \begin{smallmatrix}{D}_\mathbf{p}^K(t-t^{\prime})
    &{D}_\mathbf{p}^R(t-t^{\prime}) \\ {D}_\mathbf{p}^A(t-t^{\prime})&0 \end{smallmatrix}
\right)$,
where we have written $\mathbf{p}=(\mathbf{p},p_z)$.  In frequency space
the retarded and \index{Keldysh}Keldysh \index{Green function}Green's functions are given by
$ {D}_\mathbf{p}^{R}(\omega) = [\hbar\omega-\hbar\omega^{\zeta}_\mathbf{p}
+ i\delta]^{-1}, {D}_\mathbf{p}^K(\omega) = (-2\pi i)
F_{\Xi}(\omega)
\delta(\omega-\omega^{\zeta}_\mathbf{p}). $
Here $F_{\Xi}(E)=1 + 2n_\Xi(E)$ where $n_\Xi(E)$ is the
occupation function for the bath modes, which can have any form. For
our situation, the decay bath is empty.

Before proceeding further, we make some simplifying assumptions about
the baths.  We assume the bath frequencies $\omega^{\zeta}_{\mathbf{p},p_z}$
form a dense spectrum, and the coupling constant
$\zeta_{\mathbf{p},p_z}=\zeta(\omega^{\zeta}_{\mathbf{p},p_z})$ is a smooth
function.  We may then replace summation over bath modes by
integration.  Then, taking the \index{pumping!and decay bath}bath density of states and
$\zeta(\omega)$ to be frequency independent, we have:
\begin{equation}
S^{\mathrm{decay}}_{\mathrm{bath}}=
\int_{-\infty}^\infty d\omega \sum_{\mathbf{p}}
\Psi^{*}_\mathbf{p}(\omega) 
\left(
  \begin{array}{cc}
    0 &  - i \kappa_c  \\
    i \kappa_c & 2i\kappa_c F_{\Xi}(\hbar\omega)
  \end{array} 
\right) 
\Psi_\mathbf{p}(\omega).
\end{equation}
We follow an analogous procedure for the pumping baths (see \cite{szymanska_keeling_07} for details).

\subsection{Integration over Fermionic Fields}
\label{sec:szymanska-integr-over-ferm}

%AGAIN, THESE DO NOT APPEAR TO BE FIELDS.

After integrating over the \index{pumping!and decay bath}bath degrees of freedom the full action $S$ is:
\begin{multline}
  S= 
  \iint_{-\infty}^\infty  dt dt^{\prime} 
  \left[
    \sum_{j}
    \Lambda^{*}_j(t) G_{j}^{-1}(t,t^{\prime})
    \Lambda_{j}(t^{\prime})
    +
    \sum_\mathbf{p} 
    \Psi^{*}_\mathbf{p}(t) 
    D^{-1}_{(0),\mathbf{p}}(t,t^\prime)
    \Psi_\mathbf{p}(t^{\prime})
  \right],
 \\
  \label{eq:szymanska-bare-phot-gf}
  \mbox{where } D^{-1}_{(0),\mathbf{p}}(t,t^\prime) = 
    \left(
      \begin{array}{cc}
        0 & i\hbar\partial_t -\hbar\omega_\mathbf{p} - i\kappa_c \\
        i\hbar\partial_t -\hbar\omega_\mathbf{p} +i\kappa_c & 2i\kappa_c F_{\Xi}(t-t^{\prime})
      \end{array} 
    \right). 
\end{multline}
To specify the \index{exciton}exciton \index{Green function}Green's function $G_j$ we first
introduce the abbreviations
$
  \lambda_{\textrm{cl}}(t)=\sum_{\mathbf{p}}
  g_{j}\Psi_{\mathbf{p},cl}(t)/\sqrt{2}
$
and
$
  \lambda_{\textrm{q}}(t)=\sum_{\mathbf{p}}
  g_{j}\Psi_{\mathbf{p},q}(t)/\sqrt{2}
$
so that:
\begin{equation}
  \label{eq:szymanska-Gtotal}
  G^{-1}_j
  =
  \left(
    \begin{array}{cccc}
      0 & -\lambda_{\textrm{q}}(t) & i \hbar\partial_t - {\epsilon_j} - i \gamma_x & - \lambda_{\textrm{cl}}(t) \\
      -\lambda^{*}_{\textrm{q}}(t) & 0 & - \lambda^{*}_{\textrm{cl}}(t) & i \hbar\partial_t + {\epsilon_j} - i \gamma_x \\
      i \hbar\partial_t - {\epsilon_j} + i \gamma_x & - \lambda_{\textrm{cl}}(t) & 2 i \gamma_x F_D(i \hbar\partial_t) & -\lambda_{\textrm{q}}(t) 
      \\
      - \lambda^{*}_{\textrm{cl}}(t) & i \hbar\partial_t + {\epsilon_j} + i \gamma_x &  -\lambda^{*}_{\textrm{q}}(t) & 2 i \gamma_x F_C(i \hbar\partial_t)
    \end{array}
  \right),
\end{equation}
where $F_{C,D}(E) = 1 - 2n_{C,D}(E)$ with $n_{C,D}(E)$ the pumping
\index{pumping!and decay bath}bath occupation functions.  As the occupation functions of all baths
appear in this action, they compete to set the occupation function of
the \index{polariton}polaritons.  This \index{non-equilibrium!quantum field theory}non-equilibrium action thus combines strong
\index{exciton-photon}\index{exciton}exciton-photon\index{interaction(s)!exciton-photon} coupling with the effects of dissipation due to the
open nature of the system.  The action is quadratic also in the
fermionic fields $\Lambda_j$, so we can also integrate over these
fields to get the \index{effective action}effective action for the photon field alone:
\begin{equation}
  S = 
  -i\sum_{j}\tr  \left\{ \ln 
  G_{j}^{-1} \right\} +
  \iint_{-\infty}^\infty \!\!\! dt dt^{\prime} \sum_{\mathbf{p}}
  \Psi^{*}_\mathbf{p}(t) D^{-1}_{(0),\mathbf{p}}(t,t^\prime) \Psi_\mathbf{p}(t^{\prime}). 
  \label{eq:szymanska-totalS}
\end{equation}
As yet, we have made no assumption about what form $\Psi_{\mathbf{p}}(t)$
takes, however, since $\tr \{ \ln G_{j}^{-1}\}$ involves
$\Psi_{\mathbf{p}}(t)$, this \index{effective action}effective action is nonlinear, so to proceed
further analytically, some expansion or approximation scheme is
required.  Section~\ref{sec:szymanska-mean-field-condition} therefore discusses
the \index{mean-field theory}mean-field theory of this model, and how it relates to laser
theory as well as equilibrium results.

\section{Mean-Field Condition for a Coherent State}
\label{sec:szymanska-mean-field-condition}

The \index{mean-field theory}mean-field theory of the non-equilibrium system\index{non-equilibrium!system} describes a
self-consistent \index{non-equilibrium!steady state}steady state, which may be found by evaluating the
saddle point of $S$ with respect to photon field, $ \delta S/ \delta
\Psi^{*}_{\mathbf{p},cl} =\delta S/\delta \Psi^{*}_{\mathbf{p},q}=0$.
The first equation is satisfied if the quantum component vanishes,
$\Psi_{\mathbf{p},q}=0$.  For the classical component, we write
$\Psi_{\mathbf{p},cl} = \sqrt{2} \phi_{\mathbf{p}}$, so
$\phi_{\mathbf{p}}$ corresponds to the expectation of photon
annihilation.  If condensed, we consider the ansatz $
\phi_{\mathbf{p}} = \phi_0 \exp(-i \mu_S t) \delta_{\mathbf{p},0}$,
controlled by the parameters $\phi_0, \mu_S$.  For this ansatz to
satisfy the saddle point equation, one requires
\begin{equation}
  \label{eq:szymanska-SP1}
  (\hbar\omega_0 - \mu_S - i \kappa_c) \phi_0  = \frac{i}{2} \int \frac{d
    \nu}{2 \pi} G^K_{c^\dagger_j d^{\vphantom{\dagger}}_j}(\nu). 
\end{equation}
Putting the $G^K_{c^\dagger d}$ component of Eq.~(\ref{eq:szymanska-Gtotal}) into
Eq.~(\ref{eq:szymanska-SP1}) and defining $E_j^2 = (\epsilon_j - \mu_s/2)^2 + g_{j}^2\phi^2_0$ 
we have the saddle point (mean-field) equation:
\begin{multline}
  \label{eq:szymanska-SP2}
  (\hbar \omega_0 - \mu_S - i \kappa_c) \phi_0  =   
  \sum_j g^2_j \phi_0 \gamma_x  \\ \times\int \frac{d\nu}{2\pi} 
  \frac{[F_D(\nu)+F_C(\nu)]\nu + [F_D(\nu)-F_C(\nu)](\epsilon_j-\mu_S/2+i\gamma_x)}{%
     [(\nu-E_j)^2+\gamma_x^2][(\nu+E_j)^2 + \gamma_x^2]}.
\end{multline}

As noted above, the pumping bath\index{pumping!and decay bath} occupations are imposed by choice,
and we choose these to model a thermalised reservoir of high
energy \index{exciton}excitons, with a population set by the strength of pumping.  In
order to obey on average the constraint that we consider two-level
systems, we take $n_C(\nu)+n_D(\nu)=1$.  Introducing parameters
$\mu_B, \beta_B$ to describe the occupation and temperature of this
\index{exciton}exciton reservoir, we thus define:
\begin{equation}
  \label{eq:szymanska-FAB}
  F_{C,D}(\nu)=\tanh\left[
    \frac{\beta_B}{2} \left(
      \nu \pm \frac{\mu_B-\mu_S}{2} 
    \right)
  \right].
\end{equation}
($\mu_S$ appears here via a \index{gauge! transformation}gauge transform required to remove
explicit time dependence from the \index{effective action}effective action).  If there were no
\index{exciton-photon}\index{exciton}exciton-photon\index{interaction(s)!exciton-photon} coupling the \index{exciton}excitonic two-level systems would
be thermally occupied, i.e.
 $ \langle
d_j^{\dagger}d^{\vphantom{\dagger}}_j-c_j^{\dagger}c_j^{\vphantom{\dagger}} \rangle =
-\tanh [ \beta_B (
    \epsilon_j-\mu_B/2+\mu_S/2)/2]$.

As anticipated above, Eq.~(\ref{eq:szymanska-SP2}) is rather general,
encompassing limits that correspond both to the equilibrium
gap-equation for our model \index{polariton}polariton system
(discussed in Section \ref{sec:szymanska-mean-field-theory-2}), as
well as being capable of recovering the standard laser limit
(discussed in Section \ref{sec:szymanska-mean-field-theory}).  In
addition, if one extends this approach to slowly varying condensates,
then as discussed in Section \ref{sec:szymanska-local-dens-appr}, one
may make contact with the \index{Gross--Pitaevskii equation!
  complex}complex Gross--Pitaevskii approach.

\subsection{Equilibrium Limit of Mean-Field Theory}
\label{sec:szymanska-mean-field-theory-2}

The simplest limit of the self-consistency equation,
Eq.~(\ref{eq:szymanska-SP2}), is the \index{equilibrium!thermal}thermal equilibrium limit, which
corresponds to taking $\gamma_x, \kappa_c \to 0$.  In taking this limit, it
is necessary to send $\kappa_c \to 0$ first and then $\gamma_x \to 0$.
This is because the self-consistency equation contained only the
coupling of coherent photons to the decay bath\index{pumping!and decay bath}, hence the decay
bath cannot impose a non-trivial distribution on the system,
while the pumping bath can.  In order to satisfy Eq.~(\ref{eq:szymanska-SP2})
with $\kappa_c=0$, the imaginary part of the right hand side must
vanish.  The most general way to achieve this is to set $F_D(\nu) =
F_C(\nu)$, which, considering Eq.~(\ref{eq:szymanska-FAB}), implies
$\mu_S=\mu_B$. That is,  in the absence of decay, one has chemical
equilibrium between the pumping bath\index{pumping!and decay bath} and the system.  With
$\mu_S=\mu_B$, the remaining part of Eq.~(\ref{eq:szymanska-SP2}) becomes:
\begin{equation}
  (\hbar \omega_0 - \mu_B) \phi_0  =
   \sum_j g^2_j \phi_0 \gamma_x \int \frac{d\nu}{2\pi}
     \frac{2\tanh\left({\beta_B\nu}/{2} \right)\nu}{%
     [(\nu-E_j)^2+\gamma_x^2][(\nu+E_j)^2 + \gamma_x^2]}.
\end{equation}
In the limit of small $\gamma_x$, one may use that
$\lim_{\gamma_x \to 0} 2 \gamma_x/[(\nu-E_j)^2+\gamma_x^2] =
2\pi \delta(\nu-E_j)$ to find $(\hbar\omega_0 - \mu_B) \phi_0 = \sum_j
(g^2_j \phi_0/2E_j) \tanh\left(\beta_B E_j/2 \right).$
This is the equilibrium mean-field
theory~\cite{eastham_littlewood_00,keeling_eastham_04,marchetti_keeling_06}
of the system Hamiltonian\index{Hamiltonian!polariton gas} introduced above.

\subsection{High Temperature Limit of Mean-Field Theory --- Laser}
\label{sec:szymanska-mean-field-theory}

An alternative limit to \index{equilibrium!thermal}thermal equilibrium is that of a simple
laser. This limit too can be recovered from
Eq.~(\ref{eq:szymanska-SP2}), 
in this case by taking $F_{C,D}(\nu)$ to be frequency
independent.  This frequency independence can be recovered
from Eq.~(\ref{eq:szymanska-FAB}) in the limit $T \to \infty$,
while keeping $\mu_B \propto T$ in order that the bath\index{pumping!and decay bath}
population remains fixed.  Another interpretation of this is
that infinite temperature corresponds to white \index{noise!white}noise, i.e.\ a
\index{Markov approximation}\index{Markov approximation|seealso{non-Markovian}}Markovian approximation, where the occupation of the bath modes is
frequency independent.  [In contrast, Eq.~(\ref{eq:szymanska-SP2})
  has a flat density of states of the bath\index{pumping!and decay bath}, but a \index{non-Markovian}non-Markovian,
  i.e.\ frequency dependent, occupation].

Taking $F_{C,D }(\nu)$ to be frequency independent, the integral in
Eq.~(\ref{eq:szymanska-SP2}) can then be simply evaluated by contour integration
to give:
\begin{equation}
  (\hbar\omega_0 - \mu_S - i \kappa_c) \phi_0
  =
  \sum_j g^2_j \phi_0  (F_D-F_C)
  \frac{\epsilon_j - \mu_S/2 + i\gamma_x}{4(E_j^2+\gamma_x^2)}.
\end{equation}
The term on the right hand side, describing the two-level system
polarisation, is proportional to the bath inversion $N_0 = (n_D-n_C) =
-(F_D-F_C)/2$.  In the limit $\phi_0 \to 0$, this equation recovers
the standard threshold condition for a laser~\cite{haken_book_70}.
This is clear if one restricts to $g_j=g, \epsilon_j=\epsilon$ so the
sum is replaced by a factor $n$, and one assumes resonance, $2
\epsilon = \hbar\omega_0 = \mu$, which yields: $2 \kappa_c
\gamma_x/{g^2}=n N_0=$total inversion.

\subsection{Low Density Limit: \index{Gross--Pitaevskii equation! complex}Complex Gross--Pitaevskii Equation}
\label{sec:szymanska-local-dens-appr}

Equation~(\ref{eq:szymanska-SP2}) is written for a uniform \index{non-equilibrium!steady state}steady state, but in
many cases, it is interesting to allow for solutions that vary slowly
in time and space.  To do this rigorously requires some care, but the
basic idea can be described simply: One may start by writing
Eq.~(\ref{eq:szymanska-SP2}) in the form $(\mu_S + i \kappa_c - \hbar\omega_0 )
\phi = \chi[\phi, \mu_S] \phi$, where $\chi[\phi,\mu_S]$ is a
nonlinear complex susceptibility.  If one then separates the fast and
slow time dependence $\Psi(\mathbf{r},t) = \phi(\mathbf{r},t) e^{- i \omega_0 t}$, one
may write $ ( i \hbar \partial_t + i \kappa_c - [ V(\mathbf{r}) - \hbar^2
  \nabla^2/2m ] ) \phi(\mathbf{r},t) = \chi[\phi(\mathbf{r},t)] \phi(\mathbf{r},t)$, having
introduced an external potential $V(\mathbf{r})$.  Then, by making a gradient
and Taylor expansion of the nonlinear complex susceptibility
$\chi[\phi]$, one is naturally led to a \index{Gross--Pitaevskii equation! complex}complex Gross--Pitaevskii
equation: $ i \hbar \partial_t \phi = ( - \hbar^2 \nabla^2/2m^\ast +
V_{\text{eff}}(\mathbf{r}) + U |\phi|^2 + i [\gamma_{\mathrm{net}}(\mu_B) -
  \Gamma |\phi|^2 ] ) \phi,$ where $\Gamma$ represents a nonlinearity
of the imaginary part of the susceptibility.  The dynamics of the
\index{exciton}excitons are responsible for producing an effective \index{polariton}polariton mass and effective potential.  In some cases, it may also be important to
consider the dynamics of the reservoir \index{exciton}excitons more carefully, by
introducing an extra degree of freedom to describe them.

\section{Applications: Fluctuations and Instability Towards BEC}
\label{sec:szymanska-fluct-inst-norm}

So far, we have discussed only the mean-field properties of the
\index{non-equilibrium!condensate}non-equilibrium \index{condensate!polariton}\index{polariton}polariton condensate.  We next consider
\index{fluctuations!thermal}fluctuations about this mean field, and in particular the photon
\index{Green function}Green's function.  This is important for several reasons.  Firstly,
knowledge of the \index{fluctuations!thermal}fluctuations determines whether a state is stable
(i.e.\ do fluctuations grow or decay in time).  Secondly, the
photon \index{Green function}Green's function describes the fluctuation contribution to
physical observables such as luminescence and absorption spectra.

To determine the photon \index{Green function}Green's function, one may start from
Eq.~(\ref{eq:szymanska-totalS}), and expand $\Psi = \phi + \psi$ to second order
in $\psi$.  The inverse photon \index{Green function}Green's function has two parts, one
from the bare photon action [given in Eq.~(\ref{eq:szymanska-bare-phot-gf})],
and one from expanding the trace over \index{exciton}excitons.  To determine the
\index{exciton}exciton part one may write
$G_{j}^{-1}
  =
  (G^{\mathrm{sp}}_{j})^{-1}
  +
  \delta G_{j}^{-1}$,
where $G^{\mathrm{sp}}_j$ is the fermionic \index{Green function}Green's function including
the mean-field photon field $\phi$, and $\delta{G}_{j}^{-1}$ is the
photon fluctuation part given by:
$\delta G^{-1}_j= -{g_j}( \psi^{*}_{\textrm{q}} \sigma_{-}^{dc} + \psi_{\textrm{q}}
\sigma_{+}^{dc} )\sigma_{0}^{\textsc{k}}/{\sqrt{2}} - {g_j}( \psi^{*}_{\textrm{cl}}
\sigma_{-}^{dc} + \psi_{\textrm{cl}} \sigma_{+}^{dc}
)\sigma_{1}^{\textsc{k}}/{\sqrt{2}}$ where $\sigma^{dc}$ are Pauli matrices
in the space of fermionic fields $c,d$.  The action then depends on
\begin{equation}
\label{eq:szymanska-so}
\tr   \left\{ \ln
 G_{j}^{-1} \right\}
  =
  \tr \left\{
    \ln \left[
     (G^{\mathrm{sp}}_{j})^{-1} \right]
    +
    G^{\mathrm{sp}}_{j} \delta G_{j}^{-1} 
    -\frac{1}{2}
      G^{\mathrm{sp}}_{j} \delta G_{j}^{-1}
      G^{\mathrm{sp}}_{j} \delta G_{j}^{-1}
  \right\}.
\end{equation}
The last term gives a contribution quadratic in $\psi$, which
contributes to the inverse photon \index{Green function}Green's function.

When considering the condensed state, it is necessary to allow for
anomalous correlations.  This requires writing the \index{Green function}Green's function in
a (Nambu) vector space of $(\psi_k, \psi^{*}_{-k})$, combined with
the $\pm$ space due to the \index{Keldysh}Keldysh/retarded/advanced structure. Thus,
in the condensed case, the \index{Green function}Green's function is a $4\times 4$ matrix,
while when non-condensed it is only a $2\times 2$ matrix. We begin by
considering \index{fluctuations!thermal}fluctuations in the normal state, and the nature of the
instability to the condensate, and then in
Section \ref{sec:szymanska-fluct-cond-syst} briefly discuss \index{fluctuations!thermal}fluctuations in the
condensed state.

\subsection{Normal State \index{Green function}Green's Functions and BEC Instability}
\label{sec:szymanska-results-polar-model}

In the normal state, the spectrum and its occupation are determined by
three real functions, the real and imaginary parts of the inverse
retarded \index{Green function}Green's function $[D^R(\mathbf{p},\omega)]^{-1} = A(\mathbf{p},\omega)
+ i B(\omega)$ and the inverse \index{Keldysh}Keldysh \index{Green function}Green's function
$[D^{-1}(\omega)]^K=i C(\omega)$.  These functions can be read
off from the fluctuation action.  Once these functions are known, one
may invert these expressions to find the retarded and \index{Keldysh}Keldysh \index{Green function}Green's
functions, and thus determine the density of states $\rho(\mathbf{p},\omega) =
-2 \im [D^R(\mathbf{p},\omega)]$ and occupation of the modes
$2n_{\psi}(\omega)+1 = i D^K(\omega)/\rho(\omega)$:
\begin{equation}
  \rho(\mathbf{p},\omega) = \frac{ 2B(\omega)}{A(\mathbf{p},\omega)^2+B(\omega)^2},
  \qquad
  n_\psi(\omega) = \frac{1}{2}\left[ 
    \frac{C(\omega)}{2B(\omega)}
    - 1
  \right].
\end{equation}
In terms of these, physical observables can be found, such as the
luminescence $\mathcal{L}(\mathbf{p},\omega) = \rho(\mathbf{p},\omega) n_\psi(\omega)$.
The roles of $A(\mathbf{p},\omega),B(\omega),C(\omega)$ can be understood by
considering the contribution from the bare photon action. In this case
$A(\mathbf{p},\omega)= \hbar\omega-\hbar\omega_p$ determines the locations of
the normal modes, $B(\omega) = \kappa_c$ gives the linewidth of these
modes and $C(\omega) = 2 \kappa_c [2 n_\Xi(\omega) + 1]$ describes
their occupation.  Including the effect of the \index{exciton}excitons, the zeros of
$A(\mathbf{p},\omega)$ now describe \index{polariton}polaritons, rather than bare photons, and
in addition $B(\omega)$ is no longer constant, hence it plays a second
role: If $B(\omega)$ vanishes at some $\hbar\omega=\mu_{\text{eff}}$ then
this causes the occupation $n_\psi(\mu_{\text{eff}})$ to
diverge. However, as long as $A(\mathbf{p},\mu_{\text{eff}}/\hbar)$ does not vanish,
the density of states will be zero at $\mu_{\text{eff}}$, so the
luminescence will remain finite.

A diverging occupation and vanishing density of states is exactly what
would, in equilibrium, occur at the \index{chemical potential}chemical potential, hence the
identification $\mu_{\text{eff}}$.  Although the non-equilibrium
system may be far from thermal, the emergence
of zeros of $B(\omega)$ thus still describes an effective chemical
potential.   To ensure $\forall_{\mathbf{p}} A(\mathbf{p},\mu_{\text{eff}}/\hbar) \neq
0$, it is necessary that $\mu_{\text{eff}}$ is below any of the zeros
of $A(\mathbf{p},\omega)$, i.e.\ the \index{chemical potential}chemical potential is below all the
\index{polariton}polariton modes, as  expected in the normal state.  We next
discuss the instability as $\mu_{\text{eff}}$ approaches the bottom of
the \index{polariton}polariton spectrum.
If $\mu_{\text{eff}}$ is near the bottom of the \index{polariton}polariton spectrum, we
can expand $A(\mathbf{p},\omega), B(\omega)$ near their simultaneous zero,
i.e.\ $A(\mathbf{p},\omega) = \alpha(\hbar\omega - \xi_{\mathbf{p}})$, and $B(\omega) =
\beta(\hbar\omega - \mu_{\mathrm{eff}})$.  One may then find where the
actual complex poles $\omega^{(D^R)}_{\mathbf{p}}$ of the retarded \index{Green function}Green's
function occur: $ \hbar\omega^{(D^R)}_{\mathbf{p}} = [(\alpha^2 \xi_{\mathbf{p}} + \beta^2
\mu_{\mathrm{eff}}) + i \alpha\beta (\mu_{\mathrm{eff}} -
\xi_{\mathbf{p}})][\alpha^2+\beta^2]^{-1}$.  These poles determine the response
to a small perturbation; thus, for perturbations to decay the poles
must have a negative imaginary part.  Hence, if $\mu_{\mathrm{eff}} >
\xi_{\mathbf{p}}$, then perturbations at that $\mathbf{p}$ grow and the normal
state is unstable. One may also show that the point where
$\mu_{\mathrm{eff}} = \xi_0$, coincides with the first point where it
is possible to satisfy the mean-field equation, Eq.~(\ref{eq:szymanska-SP2})
with $\phi=0$ and $\mu_S = \mu_{\mathrm{eff}} = \xi_0$.

One can now understand the behaviour of the non-equilibrium system\index{non-equilibrium!system} as
pumping (and hence $\mu_B$) increases: At very weak pumping, where
$\mu_B$ is large and negative, $B(\omega)$ is always positive
(i.e.\ decay dominates over gain), and no $\mu_{\text{eff}}$ exists.
As $\mu_B$ increases, a region of negative $B(\omega)$ develops, and
the boundaries of this region define $\mu_{\text{eff}}$ as discussed
above.  As long as $\mu_{\mathrm{eff}} < \xi_0$ the normal state
remains stable. At the critical pumping power, $\mu_{\text{eff}}$ then
reaches the lower \index{polariton}polariton mode at $\mathbf{p}=0$, the normal state becomes
marginally stable, and the mean-field equation can be satisfied.
Beyond this point, the normal state would be unstable, but the
condensed solution is now possible (and can be shown to be stable).

\begin{figure}
  \centering
  \includegraphics[width=11cm]{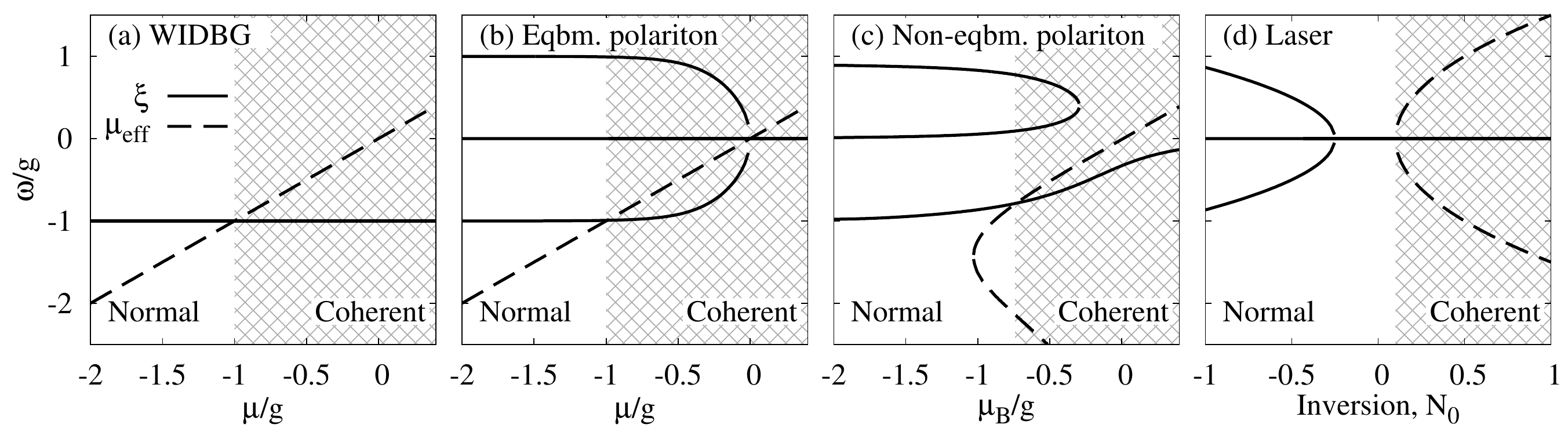}
  \caption{Trajectories of zeros of $A(\omega,\mathbf{p}=0), B(\omega)$
    (i.e.\ normal modes $\xi_0$ and effective \index{chemical potential}chemical potential
    $\mu_{\text{eff}}$) for: (a) equilibrium weakly interacting dilute
    Bose gas (WIDBG); (b) equilibrium polariton condensate; (c)
    \index{non-equilibrium!condensate}non-equilibrium polariton condensate; (d) Maxwell-Bloch laser.}
  \label{szymanska_fig1}
\end{figure}

Fig.~\ref{szymanska_fig1}(c) shows the evolution of $\mu_{\text{eff}}$
and $\xi$ for the \index{non-equilibrium!condensate}non-equilibrium \index{condensate!polariton}\index{polariton}polariton condensate.  For
comparison, Fig.~\ref{szymanska_fig1}(b) shows the behaviour of the
system Hamiltonian in \index{equilibrium!thermal}thermal equilibrium and
Fig.~\ref{szymanska_fig1}(a) that of weakly interacting and dilute
Bose gas.  One may note that despite the absence of a thermal
distribution in Fig.~\ref{szymanska_fig1}(c), the scenarios of normal
state instability in these two figures are very similar, and would
remain the same as long as the distribution function develops a
divergence while the \index{polariton}polariton system remains in strong coupling.  In
the next section, we discuss a case where the instability is somewhat
different, that of the simple laser discussed in Section
\ref{sec:szymanska-mean-field-theory} and shown in
Fig.~\ref{szymanska_fig1}(d).

\subsection{Normal-State Instability for a Simple Laser }
\label{sec:szymanska-norm-state-inst}

As in Section \ref{sec:szymanska-mean-field-theory}, one may contrast the behaviour
of the \index{non-equilibrium!condensate}non-equilibrium condensate to that of a simple laser described
by the \index{Maxwell-Bloch equations}Maxwell-Bloch equations \cite{haken_book_70}, which corresponds to
the high temperature (white \index{noise!white}noise) limit of our problem. As the above
analysis concerns the retarded \index{Green function}Green's function, we must define this
function for the \index{Maxwell-Bloch equations}Maxwell-Bloch equations.  The retarded \index{Green function}Green's
function describes the \index{linear response}linear response of the system to an applied
field, and so if one introduces a field $F e^{-i\omega t}$ coupled to
the photon field, then one has by definition
$\psi(t)=iD^R(\omega) F e^{-i\omega t}$.  One then finds:
\begin{equation}
  \label{eq:szymanska-Dlaser}
  [D^R(\omega)]^{-1} = A(\omega) + i B(\omega) = \hbar\omega - \hbar \omega_0 + i \kappa_c 
  + \sum_j\frac{g_j^2 N_0}{\hbar\omega - 2  \epsilon_j + i 2 \gamma_x}.
\end{equation}
For this form, $B(\omega)$ can only become negative if $N_0$ is
sufficiently large.  Restricting as in
Section \ref{sec:szymanska-mean-field-theory} to $g_j=g,
\epsilon_j=\epsilon=\hbar\omega_0/2$ one finds the requirement for
gain is $g^2 n N_0 > 2 \kappa_c \gamma_x$, which is again the laser
threshold condition.  In this same restricted case, the zeros of
$A(\omega)$ behave as follows: A solution $\xi=0$ always exists, and
if $N_0 < -4 \gamma_x^2/g^2{n}$, an extra pair of roots exist.  The
evolution of these zeros is shown in Fig.~\ref{szymanska_fig1}(d).
One may note that for the \index{Maxwell-Bloch equations}Maxwell-Bloch equations, strong-coupling
(i.e.\ splitting of the modes $\xi$) collapses before condensation
(lasing) occurs, while for the polariton condensate\index{condensate!polariton}, condensation
occurs while still  strongly coupled.

\subsection{Fluctuations of the Condensed System}
\label{sec:szymanska-fluct-cond-syst}

As noted earlier, if condensed, the \index{Green function}Green's function is a $4 \times
4$ matrix, so the derivation of the spectrum becomes more complicated,
however the essential features can be explained by general arguments.
The following discussion is thus based on symmetry arguments.
(The full derivation matches these results~\cite{szymanska_keeling_07}).
The form of the
inverse retarded \index{Green function}Green's function is constrained by the following
requirements: there must be symmetry under $\mathbf{p} \to -\mathbf{p}$; the modes must
in general have a finite linewidth; however at $\mathbf{p} \to 0$, there must
be a mode with vanishing frequency and vanishing linewidth
corresponding to global phase rotations of the condensate.  These
three considerations determine the leading order behaviour of
$D^R(\mathbf{p},\omega)$ for small $\omega, \mathbf{p}$.  Using these ideas, one may
then write:
\begin{equation}
  D^R(\mathbf{p},\omega) = \frac{C}{\mathrm{det}([D^R]^{-1})}
  =
  \frac{C}{\omega^2 + 2 i \omega x - c^2 p^2}.
\end{equation}
The parameters $x,c$ describe the linewidth and sound velocity.

>From this form of $D^R(\mathbf{p},\omega)$, one finds the poles are
given by $\omega^{(D^R)}_{\mathbf{p}} = - i x \pm i \sqrt{ x^2 - c^2
  p^2}$.  At long wavelengths, these are diffusive (only an imaginary
part exists), and only above a critical momentum does a real part
emerge.  Given the generality of the argument leading to this result,
it is unsurprising to find the same structure emerges from other
approaches, see e.g.~\cite{wouters_carusotto_07a}.  Similar results
also occur for the case of a parametrically pumped
\index{polariton}polariton system~\cite{wouters_carusotto_07b} .  The
absence of a linear dispersion of energy vs momentum in the condensed
state affects some aspects of \index{superfluidity}superfluidity in
this non-equilibrium system\index{non-equilibrium!system}, however
there are also aspects of superfluid behaviour that
survive\cite{wouters_carusotto_10,keeling_11}

\section{Connection to Other Approaches}

The language of the \index{Keldysh}Keldysh \index{path
  integral}\index{path integral|seealso{functional, integral}} path
integral, and the \index{Keldysh}Keldysh Green's functions provide a
natural bridge to many other approaches that have been used to model
\index{non-equilibrium!condensate}non-equilibrium
\index{polariton}polariton\index{condensate!polariton} condensation.
We have already discussed above the connection between the
\index{mean-field theory}mean-field theory, i.e.\ the saddle-point of
the \index{Keldysh}Keldysh action, and the \index{Gross--Pitaevskii
  equation! complex}complex Gross--Pitaevskii equation
\cite{wouters_carusotto_07a,wouters_carusotto_08,keeling_berloff_08}.
In order to go beyond \index{mean-field theory}mean-field theory, the
approach discussed in this chapter makes use of
\index{Keldysh}Keldysh/retarded/advanced \index{Green function}Green's
functions to describe both the occupation of a mode, and the density
of states.  These \index{Green function}Green's functions can
naturally be related to the one particle \index{density!matrix}density
matrix $\rho(\mathbf{r},\mathbf{r}^\prime,t) = \langle
\hat\psi^\dagger(\mathbf{r},t)
\hat\psi^{\vphantom{\dagger}}(t,\mathbf{r}^\prime)\rangle=i(D^K - D^R
+ D^A)(\mathbf{r},\mathbf{r}^\prime,t,t)/2$.  Direct time evolution of
the one particle \index{density!matrix}density matrix has been used to
treat \index{polariton}polaritons in zero-\cite{whittaker_eastham_09}
and \index{one-dimensional}one-dimensional
geometries~\cite{savenko_magnusson_11}, and
\index{stochastic!method}stochastic methods for simulating
\index{density!matrix}density matrix evolution have been used to
describe a number of properties of
\index{condensate!polariton}\index{polariton}polariton condensates,
see e.g.~\cite{carusotto_ciuti_05,wouters_09}.  Another
\index{stochastic!method}\index{stochastic|seealso{noise}} stochastic
approach used for \index{condensate!exciton}\index{exciton}exciton
condensation is the \index{Heisenberg--Langevin}Heisenberg--Langevin
equations~\cite{mieck_haug_02}. Such an approach again connects
naturally to the \index{Keldysh}Keldysh formalism, with the
\index{Keldysh}Keldysh self energy due to the bath corresponding
directly to the \index{noise}\index{noise|seealso{stochastic}}noise
correlator of the \index{Heisenberg--Langevin}Heisenberg--Langevin
approach, and the retarded self energy corresponding to the
dissipation term.  Finally, there is also a clear connection between
the \index{Keldysh}Keldysh \index{Green function}Green's functions and
the \index{quantum! Boltzmann equation}quantum Boltzmann equation (see
e.g.~\cite{lifshitz_pitaevskii_book_80,kadanoff_baym_book_62} for a
derivation of the \index{quantum! Boltzmann equation}quantum Boltzmann
equation from the equations of motion for the \index{Keldysh}Keldysh
\index{Green function}Green's functions).  There have been many works
using the \index{Boltzmann equation}Boltzmann equation to model
\index{condensate!kinetics}kinetics of \index{polariton}polariton
condensation
\cite{tassone_piermarocchi_97,tassone_yamamoto_99,malpuech_dicarlo_02,
  malpuech_kavokin_02,porras_ciuiti_02,doan_cao_05,doan_cao_06,doan_cao_08}.
By considering how the \index{quantum! Boltzmann equation}quantum
Boltzmann equation arises from the \index{Keldysh}Keldysh \index{Green
  function}Green's function, one may note that in order to correctly
describe the \index{coherence}coherence properties of the condensed
state, one must include anomalous retarded self energies, modifying
the \index{polariton}polariton spectrum.

\section*{Acknowledgements} 
We acknowledge financial support from EPSRC.

\bibliography{Szymanska_Keeling.bbl}

\end{document}